\documentclass[preprint2]{aastex6}
\usepackage{graphicx}

\newcommand{\msun}{\mbox{$\rm M_\odot\,$}}
\newcommand{\lsun}{\mbox{$\rm L_\odot\,$}}

\newcommand{\mum}{\mbox{$\rm \mu m\,$}}

\def\arcmin{\hbox{$^\prime$}}
\def\arcsec{\hbox{$^{\prime\prime}$}}

\watermark{}
\setwatermarkfontsize{50pt}

\begin{document}


\title{Highly variable young massive stars in ATLASGAL clumps}


\author{M.~S.~N. Kumar}
\affil{Centre for Astrophysics Research, University of Hertfordshire, Hatfield, AL10 9AB, UK \\ Instituto de Astrof\'{i}sica e Ci\^{e}ncias do Espaco, Universidade do Porto, CAUP, Rua das Estrelas, 4150-762 Porto, Portugal }

\author{ C. Contreras Pe\~{n}a}
\affil{Centre for Astrophysics Research, University of Hertfordshire, Hatfield, AL10 9AB, UK}

\author{ P.~W. Lucas}
\affil{Centre for Astrophysics Research, University of Hertfordshire, Hatfield, AL10 9AB, UK}

\and

\author{M.~A. Thompson}
\affil{Centre for Astrophysics Research, University of Hertfordshire, Hatfield, AL10 9AB, UK}



\begin{abstract}
High-amplitude variability in Young Stellar Objects (YSOs) is usually
associated with episodic accretion events. It has not been observed so
far in massive YSOs. Here, the high-amplitude variable star sample of
ContrerasPe\~{n}a et al.(2016) has been used to search for
highly-variable($\Delta$K$\ge$1\,mag) sources coinciding with dense
clumps mapped using the 850\mum continuum emission by the ATLASGAL
survey. 18 variable sources are centred on the sub-mm clump peaks, and
coincide ($<$1\arcsec) with a 24\mum point or compact ($<$10\arcsec)
source. 13 of these 18 sources can be fit by YSO models. The 13
variable YSOs(VYSO) have luminosities of $\sim$10$^3$ \lsun, an
average mass of 8~\msun and a range of ages up to 10$^6$\,yr. 11 of
these 13 VYSOs are located in the midst of infrared dark clouds. 9 of
the 13 sources have $\Delta$K$>$2\,mag, significantly higher compared
to the mean variability of the entire VVV sample. The light curves of
these objects sampled between 2010-2015 display rising, declining, or
quasi-periodic behaviour but no clear periodicity. Light-curve
analysis using Plavchan method show that the most prominent phased
signals have periods of a few hundred days. The nature and time-scale
of variations found in 6.7 Ghz methanol maser emission (MME) in
massive stars are similar to that of the VYSO light curves. We argue
that the origin of the observed variability is episodic accretion. We
suggest that the timescale of a few hundred days may represent the
frequency at which a spiralling disk feeds dense gas to the young
massive star.

\end{abstract}

\keywords{Stars:formation --- stars:massive --- 
stars:variability Herbig Ae/Be}



\section{Introduction} \label{sec:intro}
Variability studies of young low mass stars, both in the line and
continuum, have proven to be powerful tools to decipher the physics of
star formation and pre-main-sequence evolution.  Low mass Young
Stellar Objects (YSOs) displaying optical/infrared flux variability
trace a variety of phenomena \citep{carpenter01,1996AA...310...143F}
such as accretion events in the disk, magnetospheric activity
\citep{bouvier2007}, spots \citep{boubert89} and flares on the stellar
surface. High-amplitude variability is often associated with episodic
events of accretion, well-known through objects such as FUors (named
after FU Orionis) and EXors (named after EX Lupi). Hence, they are
excellent laboratories to understand the accretion phenomenon,
especially at unresolved spatial scales representing the accretion
disk and the young star.

Higher mass young stars are elusive objects that spend their brief
youth while deeply embedded in extremely dense molecular
cores. Optical photometric variability studies of Herbig Ae/Be stars
(intermediate mass YSOs) concluded that large amplitude variability is
confined to stars with spectral types later than B8 \citep{b1}. This
result should not be surprising, considering that stars more massive
than 8~\msun\, would lack a pre-main-sequence phase appearing directly
on the zero-age-main-sequence \citep{b5,b6}. In other words, optically
visible intermediate and massive stars should have already arrived on
the zero-age-main-sequence \citep{b7} with very little episodic
accretion activity. Moreover, once a massive star becomes optically
visible, the accretion rates are low \citep{hartman93} and the
accretion luminosity itself is insignificant compared to the total
luminosity. On the other hand, the infrared counterparts ({\em IRcs})
to high-mass protostellar objects ({\em HMPO}) \citep{b3,b4},
represent some sort of a pre-main-sequence phase for the massive stars
because these objects are deeply embedded in their natal dense
molecular cores, show high accretion rates, and they are detected only
in the infrared bands. Hence, they are more likely to display
variability associated with accretion episodes. However, a systematic
search for variability in such objects is still lacking. {\em IRcs} to
{\em HMPO}'s are often visible as point-sources in the near-infrared
K-band in the midst of dark clouds. The following work is an attempt
to find variability in such sources.

The Vista Variables in the Via Lactea survey
\citep[VVV,][]{2010Minniti} provides $ZYJHK_{s}$ photometry of
$\approx 560$~deg$^{2}$ of the Galactic Bulge and the adjacent
mid-plane. In addition, the survey has yielded $\approx$ 50 to 70
epochs of $K_{s}$ photometry over a period of 5 years. The VVV survey
covers much of the Galactic fourth quadrant, which is known for its
intense activity of high-mass star formation in the Milky-Way. This
virtue of the fourth quadrant has caused it to be the focus for
several other Galactic plane surveys such as ATLASGAL
\citep{schuller09} and HiGal \citep{molinari2010}, aiming to uncover
deeply embedded high-mass stars and cluster forming cores. The
ATLASGAL survey has provided an unbiased data set of dense molecular
clumps. These are massive clumps, often associated with high-mass star
formation.  Prior to the VVV study, no eruptive variable YSOs were
known with luminosities higher than a few hundred solar luminosities
\citep{audard14}. In this work we inspect the VVV sample of 816 high
amplitude infrared variable stars discovered by \citet{carlos16a} and
we isolate variables that coincide with ATLASGAL clumps in an effort
to find variable young intermediate/high-mass stars.

\section{High-amplitude variable stars within ATLASGAL clumps}

The sample of high-amplitude ($\Delta K_{s} >1$~mag) variable stars
uncovered from the VVV survey has been described by
\citet{carlos16a}. They found 816 variable stars in the 4th Galactic
quadrant ($l$=295$^{\circ}$ to 350$^{\circ}$, $b$= -1.1$^{\circ}$ to
1.0$^{\circ}$). The identification method and the selection criteria
are detailed in \citet{carlos16a}. Here, the interest is to find
variable stars that are {\em IRcs} to {\em HMPO}'s. By definition an
{\em HMPO} should be associated with a massive molecular clump,
typical of the ATLASGAL clumps. The 816 high-amplitude variable stars
were matched with a radius of 300\arcsec\, to all the peak positions
listed in the ATLASGAL web database, resulting in 39 matched
sources. To ensure that the matched variable star actually represents
a luminous (massive) source with infrared excess (young star), rather
than an unrelated foreground or background source, we used the
following constraints in the next step of filtering. The ATLASGAL data
has an angular resolution of 19.2\arcsec, and the pointing rms better
than 4\arcsec. The variable star associated with the sub-mm clump
(almost always within the peak of the 19.2\arcsec beam) must coincide
(to better than 1\arcsec ) with a 24\mum source that is either point
or compact ($<$10\arcsec) like. 18 of the 39 stars above satisfied
these criteria. Next, the spectral energy distribution (SED) of the
source should be representative of a YSO and not a stellar
photosphere. Therefore, the 1\mum--850\mum photometric measurements
for these 18 objects were assembled using any, and all available data
from the 2MASS, GLIMPSE, MIPSGAL, WISE, HiGal and the ATLASGAL
databases (See Sec.\,3 and Table\,2). The resulting SEDs were fitted
using the SED fitting tool described by \cite{rob07} which attempts to
model the observed SED with an appropriate young stellar object model,
at the same time checking if the SED can represent a main-sequence
photosphere. Five of the 18 objects were fitted by stellar
photospheres, leaving 13 VVV sources that were fit with YSO
models. These 13 variable young stellar objects (VYSOs) are listed in
Table\,1 along with their known properties.

In Table\,1, $\sigma K_s$ is the typical scatter in each magnitude bin
of K$_s$ for a given VVV tile (see \citet{carlos16a} for a full
description). This scatter is in the range 0.4-0.6\,mag as is typical
with uncorrected pipeline photometry of the VVV tiles. The choice of
$\Delta K_{s} >1$~mag cut-off \citep[both in this work and that
  of][]{carlos16a}, is a conservative choice to select objects well
above the scatter $\sigma K_s$.  This choice is the reason why only a
small number of 13 VYSOs are found in this study. An extended study is
underway to examine the general variability characteristics of
high-mass YSOs by employing the relatively accurate, paw-print
photometry that displays lower scatter levels.

\begin{figure*}
\includegraphics[width=180mm]{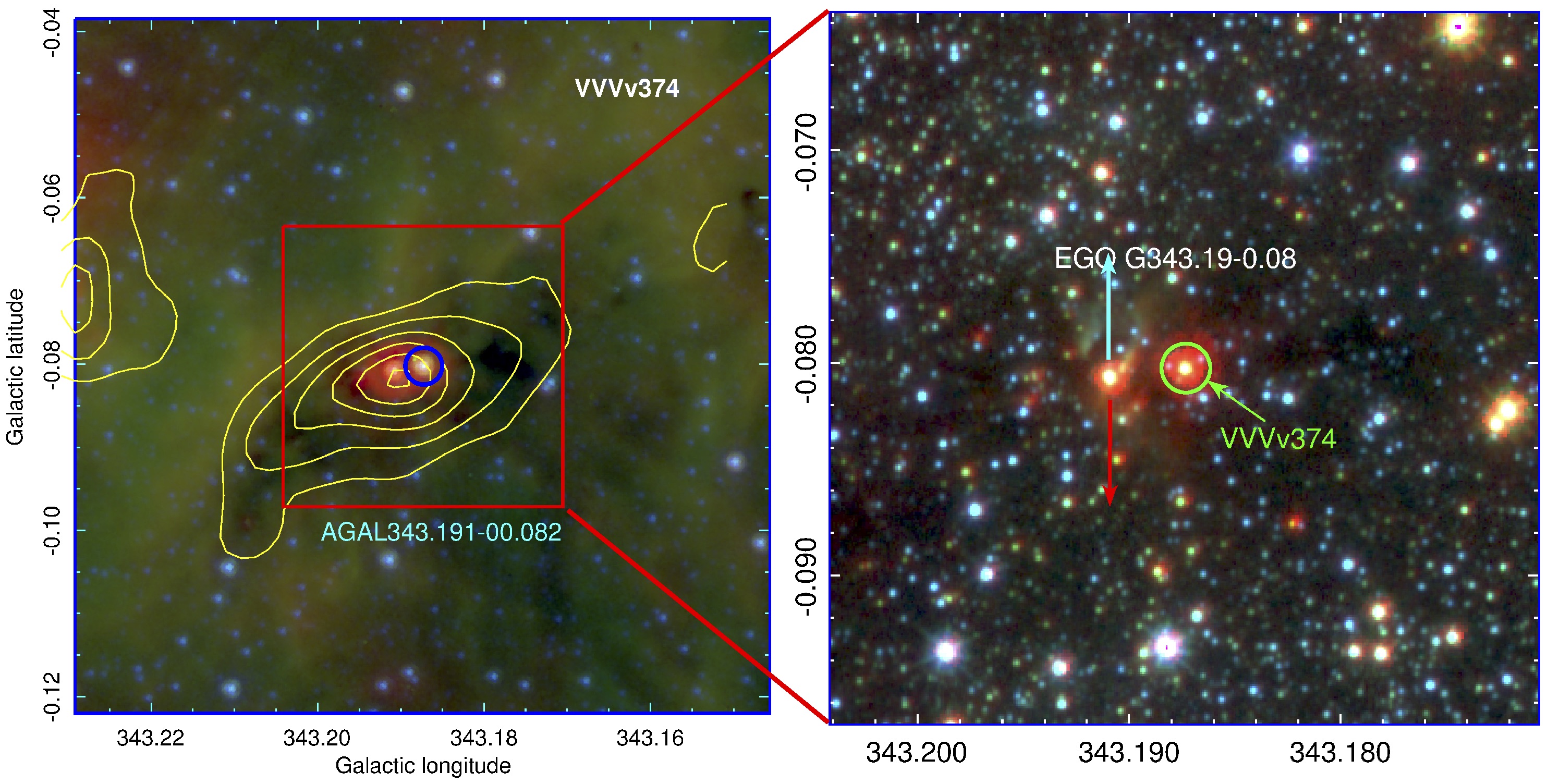}
 \caption{ The environment of VVVv374. Left Panel: Spitzer-MIPS
   24\mum, IRAC 8\mum and 3.6\mum images are coded as red, green and
   blue respectively to compose the colour image which reveal the
   infrared dark filament/cloud. Yellow contours show the 870\mum
   emission (starting at 0.1\,Jy, in steps of 0.1\,Jy) from the
   ATLASGAL data tracing the molecular clump associated with the
   infrared dark cloud. Two young stars are visible at the peak of the
   clump. The blue circle identifies the variable young star. Right
   Panel: Zoom in view of the two young stars shown with a colour
   composite of the Spitzer-IRAC 4.5\mum\, VVV K$_s$ and VVV H-band
   images coded as red, green and blue respectively. The young star
   adjacent to the variable source is associated with an extended
   green object (EGO) as shown. Blue and red arrows represent possible
   blue and red shifted lobes of the outflow associated with the
   EGO. }
\end{figure*}
\figsetstart
\figsetnum{1}
\figsettitle{VYSO colour overlays}

\figsetgrpstart
\figsetgrpnum{1.1}
\figsetgrptitle{overlays}
\figsetplot{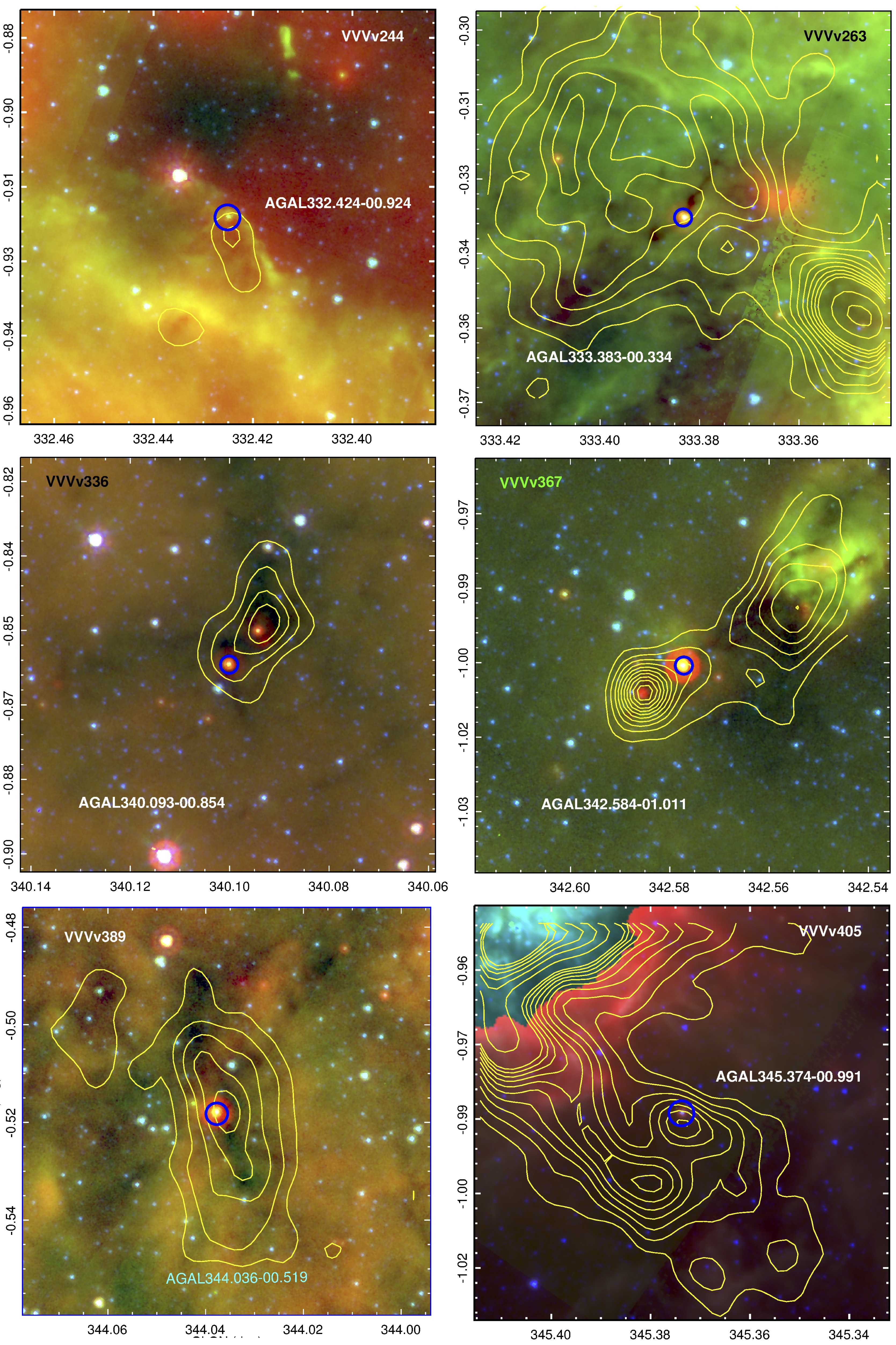}
\figsetgrpnote{Same as left panel of Fig.1}
\figsetgrpend

\figsetgrpstart
\figsetgrpnum{1.2}
\figsetgrptitle{overlays}
\figsetplot{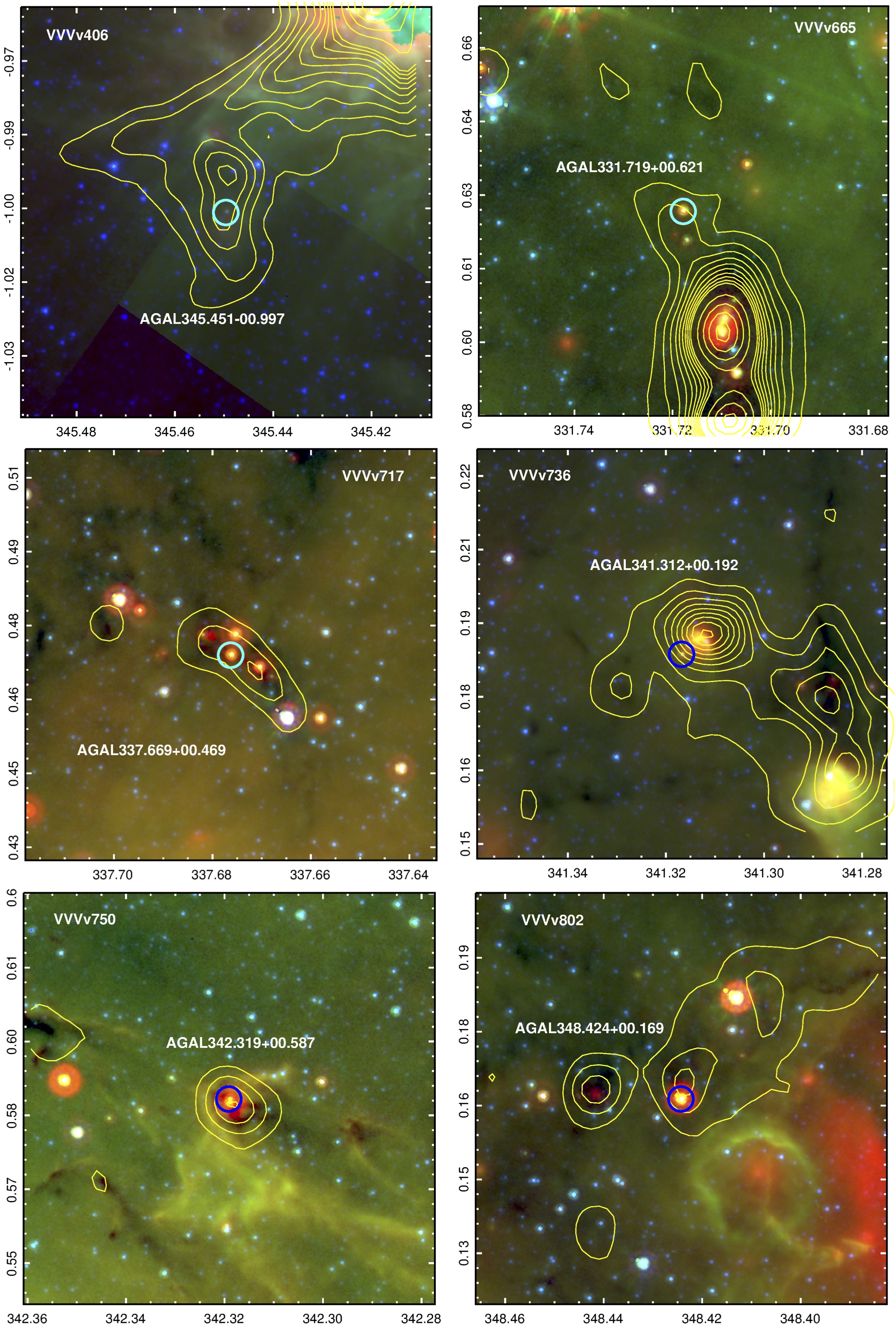}
\figsetgrpnote{Same as left panel of Fig.1}
\figsetgrpend

\figsetend

\begin{figure}
\figurenum{1}
\plotone{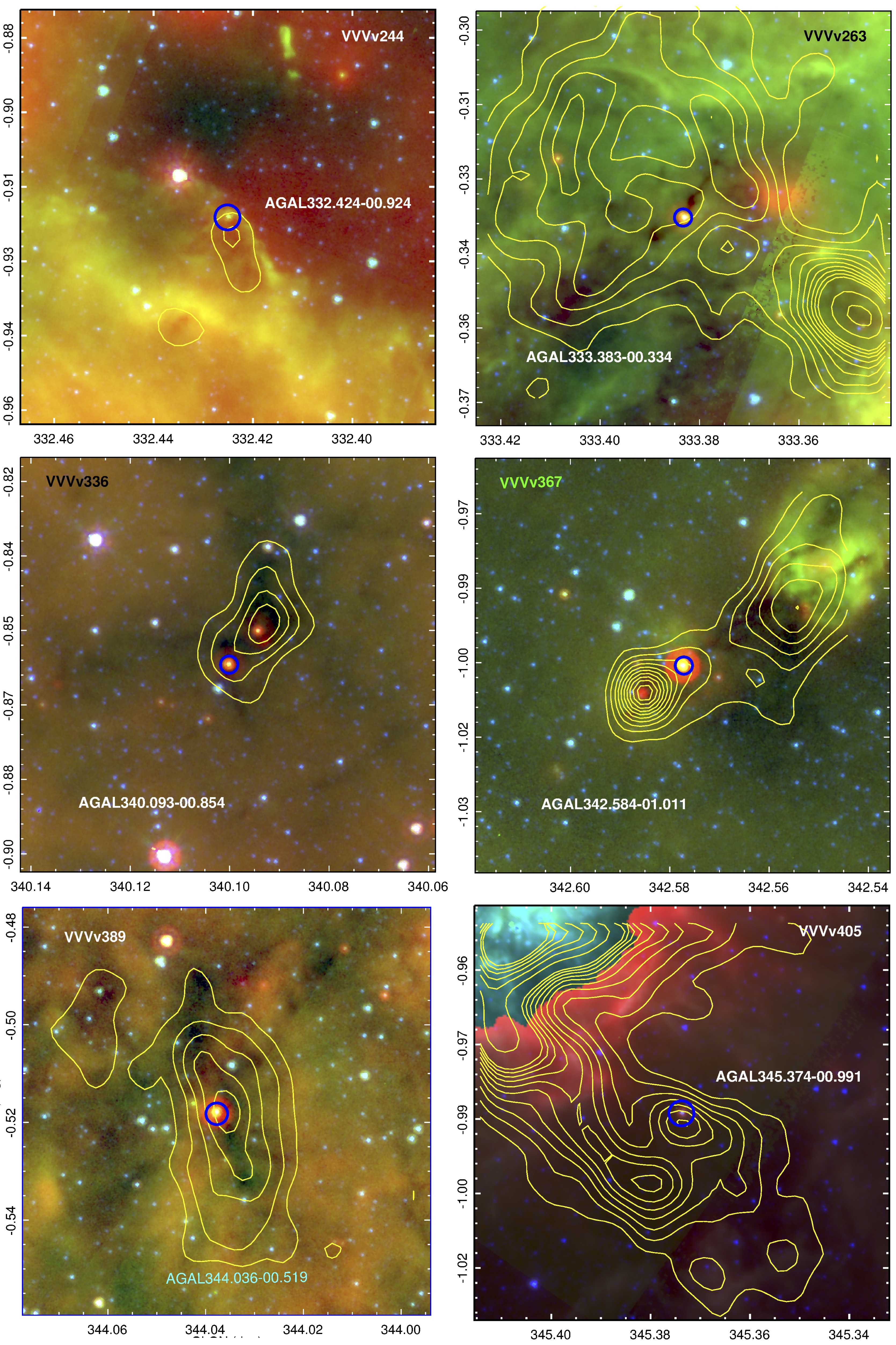}
\caption{Same as left panel of Fig.1}
\end{figure}
\begin{figure}
\figurenum{1}
\plotone{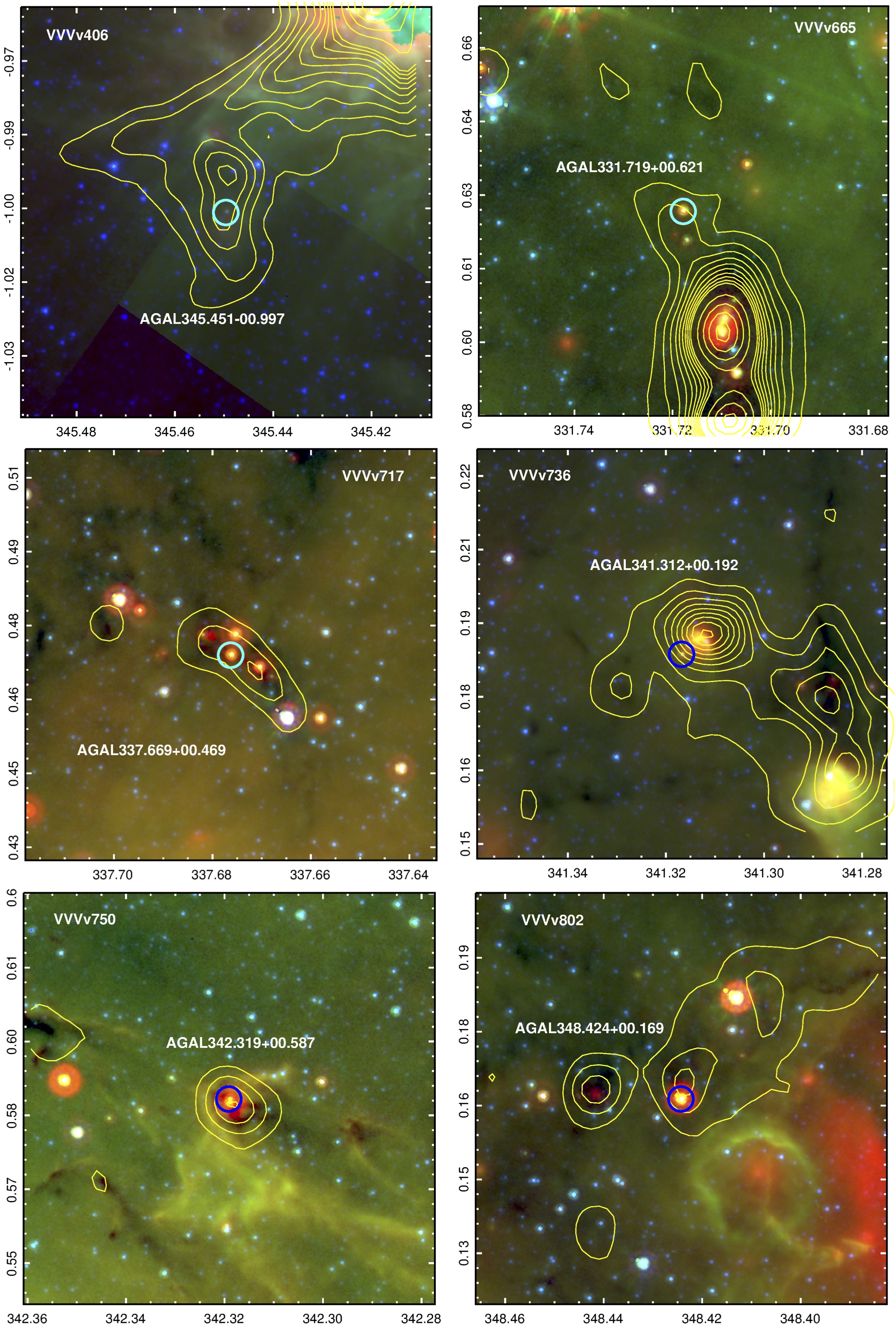}
\caption{Same as left panel of Fig.1}
\end{figure}

To visualise and examine the environment of the VYSOs, we composed
colour images using the Spitzer MIPSGAL \citep{mipsgal} and GLIMPSE
\citep{glimpse} cutouts and overlaid contours of 850\mum emission from
the ATLASGAL data. In Figure.~1(left panel) we display such an overlay
for the source VVVv374. It is an excellent example of the 13 objects
listed in Table\,1. The Spitzer MIPS 24\mum,  IRAC 8\mum, and IRAC 3.6\mum
images are red, green and blue respectively, and the yellow
contours show the ATLASGAL 850\mum emission. Similar plots for the
remaining 12 targets are shown in the supplementary Fig.\,Set\,1. In 
Figure.~1(left panel) the infrared VYSO appears as a
bright 24\mum source, embedded inside the elliptical shaped dense core
traced by the 850\mum emission contours. The dark patch in the
background image reveals the associated infrared dark cloud. In 11 of
the 13 cases, the ATLASGAL cores are associated with a catalogued
infrared dark cloud (IRDC) \citep{perful09}.  The presence of an IRDC
is an important signpost that the selected objects are indeed sites of
massive star formation. The star formation activity in VVVv374 is
further evident through the presence of an extremely green object EGO
\citep{ego}. The right panel in Figure.~1 shows this feature for
VVVv374, highlighting the bipolar outflow traced by the EGO. Here the
IRAC4.5\mum, VVV K and H band images are used as red, green and blue
respectively. The Vizier database was used to search a 30\arcsec\,
radius around each target to find known sign-posts of star formation
such as IRDCs, EGOs, Masers, and HII regions and they are listed in
the last column of Table\,1.

We note that 9/13 sources in Table 1 have amplitudes $\Delta K_{s}
>2$~mag, which is significantly higher than the mean variability of
all the YSOs in the VVV sample of \citet{carlos16a}, regardless of the
light curve classification. In that study, high amplitude was found to
be correlated with an early stage of evolution, as traced by the 2 to
22~$\mu$m spectral index. The presence of ATLASGAL cores and IRDCs
tends to support this correlation, in addition to indicating that high
amplitude variable YSOs can be massive systems.

\section{SED Fitting}

By using the SED modelling, one can qualitatively assess the
luminosity and mass of the stars of interest. The photometric data
compiled for the 13 objects is tabulated in Table\,2. It includes the
four Spitzer-IRAC bands from the GLIMPSE survey (3.6\mum, 4.5\mum,
5.8\mum, 8\mum), two bands from the  WISE survey (12\mum, 22\mum),
Spitzer-MIPS photometry at the 24\mum band \citep{rob08}, 5 bands from
the Herschel-HiGal survey (70\mum, 170\mum, 250\mum, 350\mum, 500\mum)
and the ATLASGAL 850\mum band. Whenever MIPS 24\mum photometry is
available the WISE 22\mum is ignored to ensure using the higher
quality data. Given that our targets display $>$1~mag variation in the
K band, and the above surveys are non-contemporaneous, the variations
in the near-infrared J and H bands are expected to be of similar
amplitude. Therefore they are not used to model the SEDs. In the
IRAC\,3.6\mum and 4.5\mum bands, typically 0.2-0.4\,mag variations are
reported for low mass YSOs \citep{faesi12,wolk15}. Moreover
\cite{carlos16a} typically find lower amplitude variability at these
longer wavelengths compared to K-band variation. Such variations
should not affect the overall SED and the resulting fit.

The photometric data listed in Table\,2 was fed to the SED fitting
tool \citep{rob07}. The physics of YSOs behind this tool is detailed
in \citet{rob06}. We used appropriate apertures for each band, a
classification as data or upper limit, and a range in distance and
extinction for each source. The apertures used are 4\arcsec\, for
Spitzer-IRAC and 12\arcsec\, for WISE bands and 6\arcsec diameter for
MIPS 24\mum, respectively. The full-width half maximum of the HiGal
and ATLASGAL cores were used as aperture diameters for the sub-mm
data.

The data at wavelengths shorter than (and inclusive of) 24\mum were
employed as ``data points'' (as they critically define the fitted
models), and the remaining longer wavelength data are set as ``upper
limits'' (which allow to constrain the fitted models) in the SED
fitting \citep[e.g.][]{b4}. This choice is made to minimise the
contamination from other sources enclosed within the large beams of
the long wavelength data. The SED fitting tool scales and fits YSO
models derived from a library of 200000 models. The scaling is done by
adopting an appropriate combination of extinction and distance within
a range provided by the user. Neither of these factors are known with
certainty \citep[see][]{carlos16a} for our sources, so, we used a
fixed range of A$_v$=5--50mag and d=1--13kpc for all 13
objects. Typically, the adopted choice of the SED fitter is correct
within a factor of two for both values because of the logical
combination of YSO models stored in the library. The choice of A$_v$
and the distance used by the SED fitter is an output from the fitting
results, listed in Table\,3 along with other parameters.  We note that
IRDCs in quadrant 4 are typically located at distances of $\sim$3 kpc
\citep{jackson2008}, and distances over $\sim$8~kpc are unlikely
because these dense clouds require a bright mid-infrared background in
order to be observed in silhouette. The SED fitting results are
consistent with the expected locations on the near side of the Milky
Way.

The fitting procedure is based on $\chi^2$ minimisation method and
yields a best fit model together with a range of models satisfying the
criteria ${\chi^2}_{best} - \chi^2$ per data point is less than 3. The
model parameters listed in Table\,3 are weighted means of this suite
of models, the weights being the inverse of the $\chi^2$ of each
model. In Table\,3 the column N$_{fits}$ indicate the number of models
used to compute the parameters satisfying the above criteria, and
N$_{fits2}$ are the number of models which has a non-zero value of
envelope accretion rate. The fitted YSO models for the exemplar
source VVVv374 is shown in Figure.\,2. The remaining fits are
displayed in the supplementary Fig.\,Set 2.

\begin{figure}
\includegraphics[width=78mm]{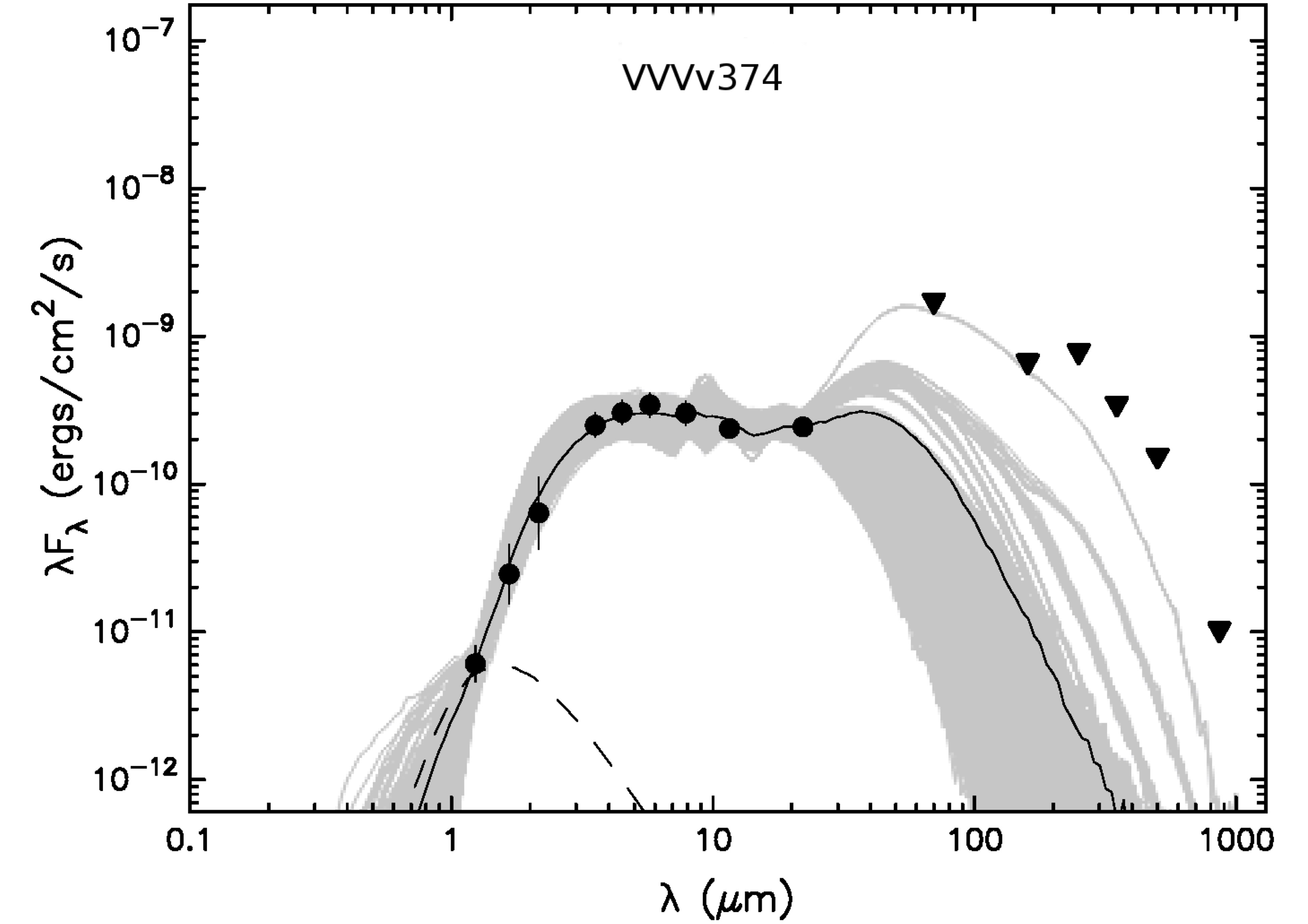}
  \caption{SED fitting results for the source VVVv374. Filled
    circles and triangles show the ``data points'' and ``upper
    limits'' respectively. Solid black line and grey lines
    show the best fit model and the models satisfying the criteria
    ${\chi^2}_{best} - \chi^2 < 3$ respectively. The dashed line
    represents the embedded photosphere of the best fit YSO model.}
\end{figure}

\figsetstart
\figsetnum{2}
\figsettitle{SED models for VYSOs}

\figsetgrpstart
\figsetgrpnum{2.1}
\figsetgrptitle{SED fitting}
\figsetplot{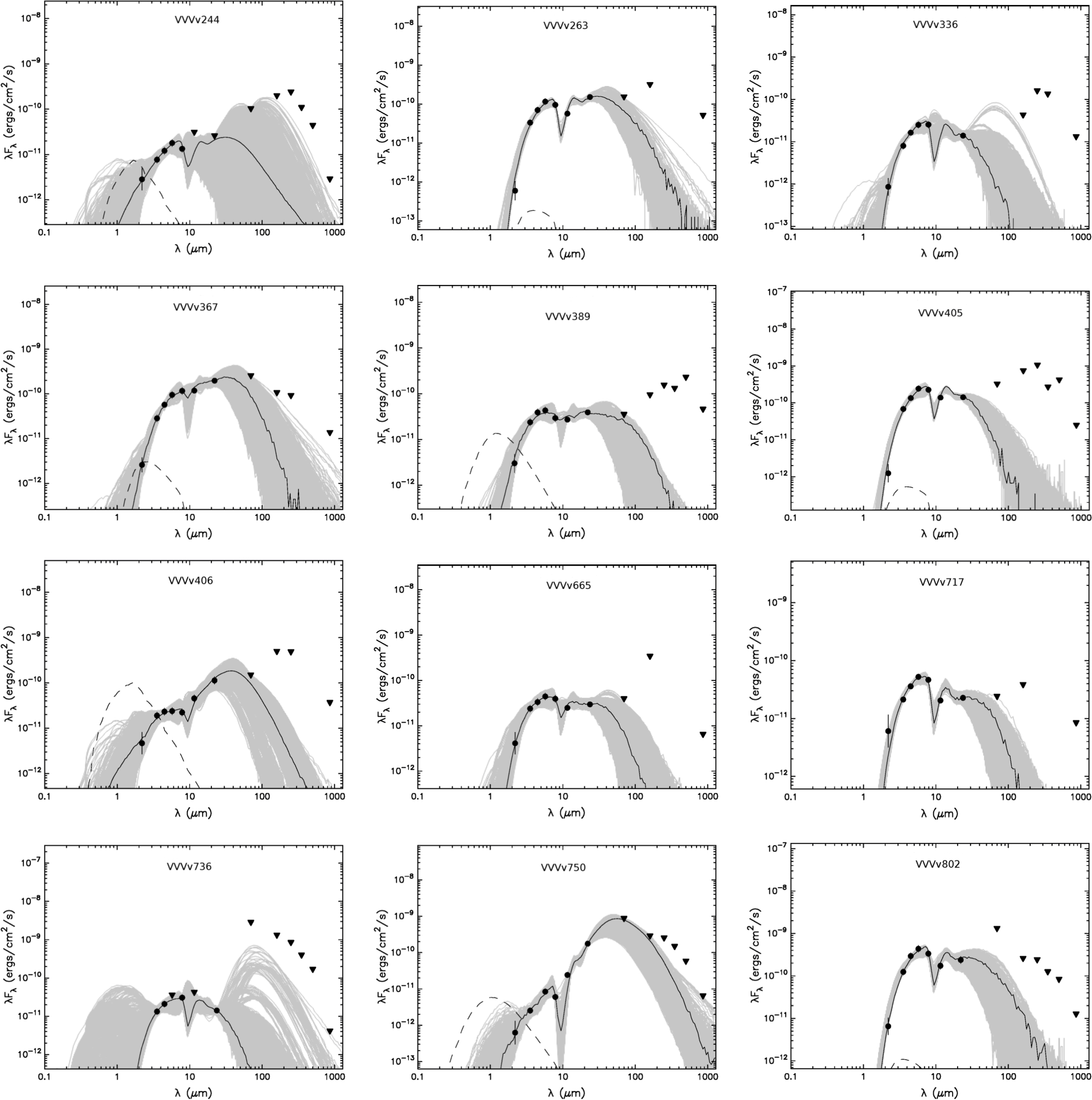}
\figsetgrpnote{Same as Fig.2}
\figsetgrpend

\figsetend

\begin{figure}
\figurenum{2}
\plotone{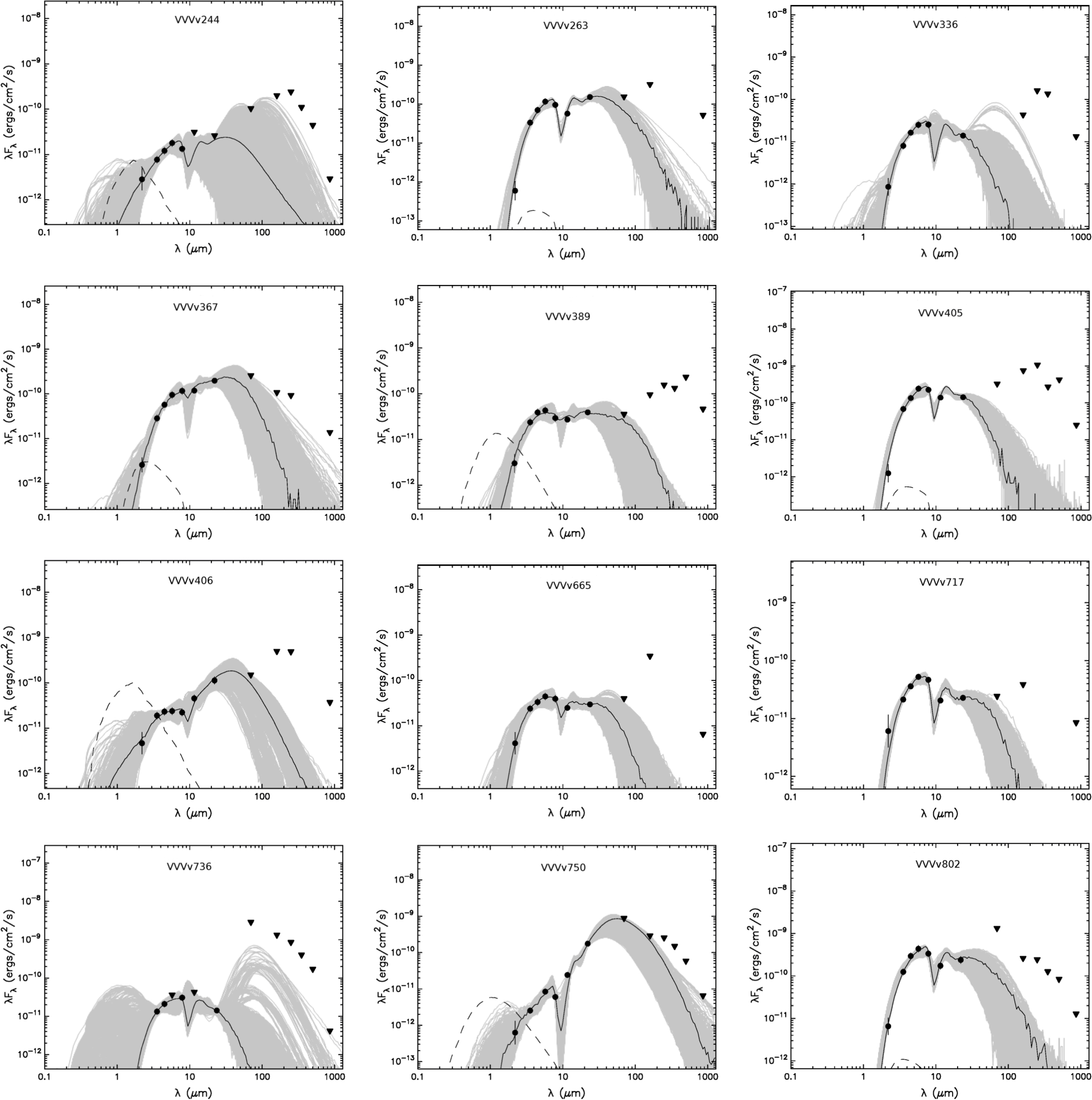}
\caption{Same as Fig.2}
\end{figure}

\subsection{SED fitting results}

The most reliable output of the SED fitting procedure is the
luminosity because they are obtained by a simple scaling of the
pre-computed radiation transfer models to match the flux peak and
shape of the observed SED by using an appropriate combination of
extinction and distance to minimise the $\chi^2$ \citep{rob07}. This
is also the reason why the extinction and distance estimates given by
the fitting procedure turn out to be correct within a factor of
two. This procedure will select optimal combinations of mass and age
to match the observed SED from an uniformly sampled parameter space
\citep[see][]{rob06} which is used to derive the temperature and
radius of the young star through evolutionary models.  From Table.~3
it can be seen that the selected high-amplitude variable stars are
best modelled as intermediate to high mass objects with an average
luminosity of 10$^{3}$~\msun\,, and mass of 8~\msun\,.

The modelled ages show that they are typically million year old
objects, with a significant level of disk and envelope accretion
rates. The number of models N$_{fits2}$ with a non-zero envelope
accretion rate are fewer compared to the total number of models,
because much of the longer wavelength data are used as upper limits,
to avoid contribution from adjacent sources. However, in 9/13 sources
the HiGal 70\mum datum corresponds to an unresolved source detected
within 5\arcsec of the VVV source. Since the Herschel/PACS beam size
is $6\arcsec \times 12$\arcsec\, at this wavelength, this strongly
implies that the 70\mum\, flux arises in the disk or envelope of
the YSO, rather than the larger scale cloud. If we were to use the
70\mum data for these 9 sources actively in the fit, instead of
upper limits, at least 6/13 sources would be fitted with a
non-zero envelope accretion rate, compared to the 2/13 sources in
Table\,3. Consequently, it appears likely that these YSOs have a
range of ages extending below one million years. We also note that
using HiGal fluxes as data points instead of upper limits would tend
to increase the total luminosity of these systems.

The 13 targets are all near-infrared visible, suggesting that they
have partially emerged from the deeply embedded phase which is often captured only at
sub-mm regime. The inferred disc accretion rates represent a mean value
of 10$^{-6}$~\msun yr$^{-1}$. It is two orders of magnitude smaller
than the high disk accretion rates measured through millimeter
observations \citep{cesa07} or from SED modelling of {\em IRcs} to
{\em HMPO}'s \citep{b4}. The fitting results can be improved by
acquiring higher spatial resolution data at longer wavelengths. The
results obtained here are merely indicative, the main inference is
that the selected targets represent YSOs of high bolometric
luminosity, therefore representing massive YSOs.

\section{Light Curves}

The VVV data obtained between 2010-2015 have been used to construct
the light curves for the 13 sources here. The light curve
from the multi-epoch photometric data for the source VVVv374 is shown
in Figure.\,3. The remaining light curves are displayed in the
supplementary Fig.\,Set 3. It can be seen that the data points are
spread out fairly uniformly along the time axis with several gaps in
between, as is typical for the survey observations. The gaps are often
larger than one week between consecutive observations but shorter,
intra-day time intervals are also sampled. The time series will nearly
always detect changes in flux that endure for several months
\citep{carlos16a}. Regular variations with periods from hours to weeks
are also likely to be detected, though one-off variations on these
short timescales are likely to be missed.  The light curve
classifications for these objects from \citet{carlos16a}, based on the
2010-2015 time series, are listed in Table.\,1.

Examining the light curves of all the 13 VYSOs, we do not find
visually obvious periodicity, except in the case of VVVv802. This
light curve was classified as LPV-Mira by \citet{carlos16a}, owing to
the excellent fit to a sine curve in the 2010-2014 time
series. However, the 2015 datum departs from the previous
trend. Moreover, the SED of this source (especially the 24\mum and
70\mum data) does not resemble those of dusty Mira variables discussed
by \citet{carlos16b}, so the Mira interpretation is disfavoured. The
observed light curve is the sum of flux variations caused by a variety
of phenomena such as accretion, spots, eruptive events, line-of-sight
extinction due to rotation of the source \citep{carpenter01}.  The
light curves of some pre-main-sequence stars are shown to have
contributions from pulsations \citep{kallinger08}. Additionally, noise
from variable seeing and weather conditions also contribute to the
observed light curve. Analysing the light curve with a periodogram can
often uncover a phased signal if it happens to be of substantial
significance.

\begin{figure}
\includegraphics[width=85mm]{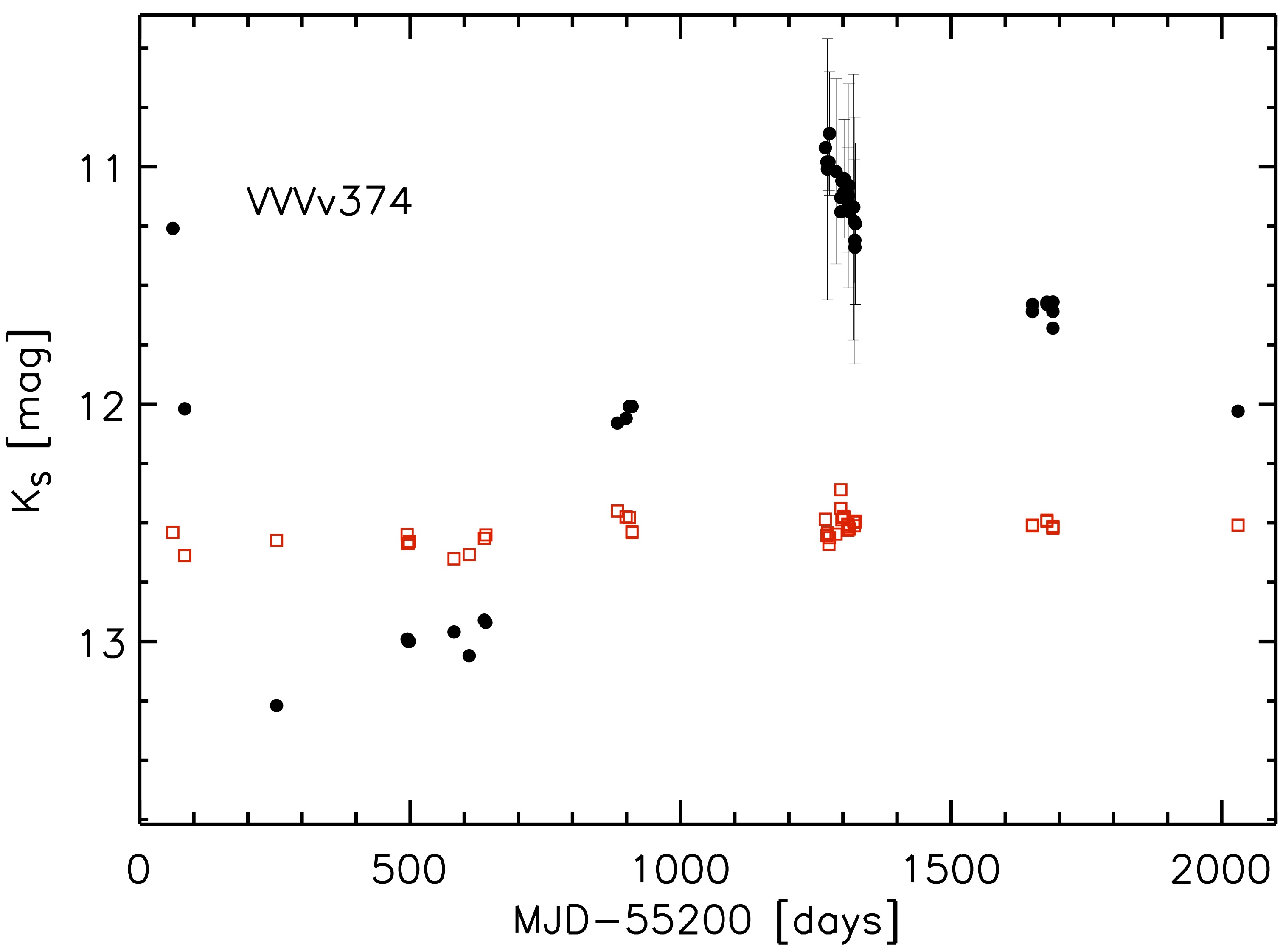}
\caption{Light curve of the source VVVv374. Red squares show the light
  curve of a non-varying control object within the same field of
  view.}
\end{figure}
\figsetstart
\figsetnum{3}
\figsettitle{SED models for VYSOs}

\figsetgrpstart
\figsetgrpnum{3.1}
\figsetgrptitle{I couldn't decipher your translation table!}
\figsetplot{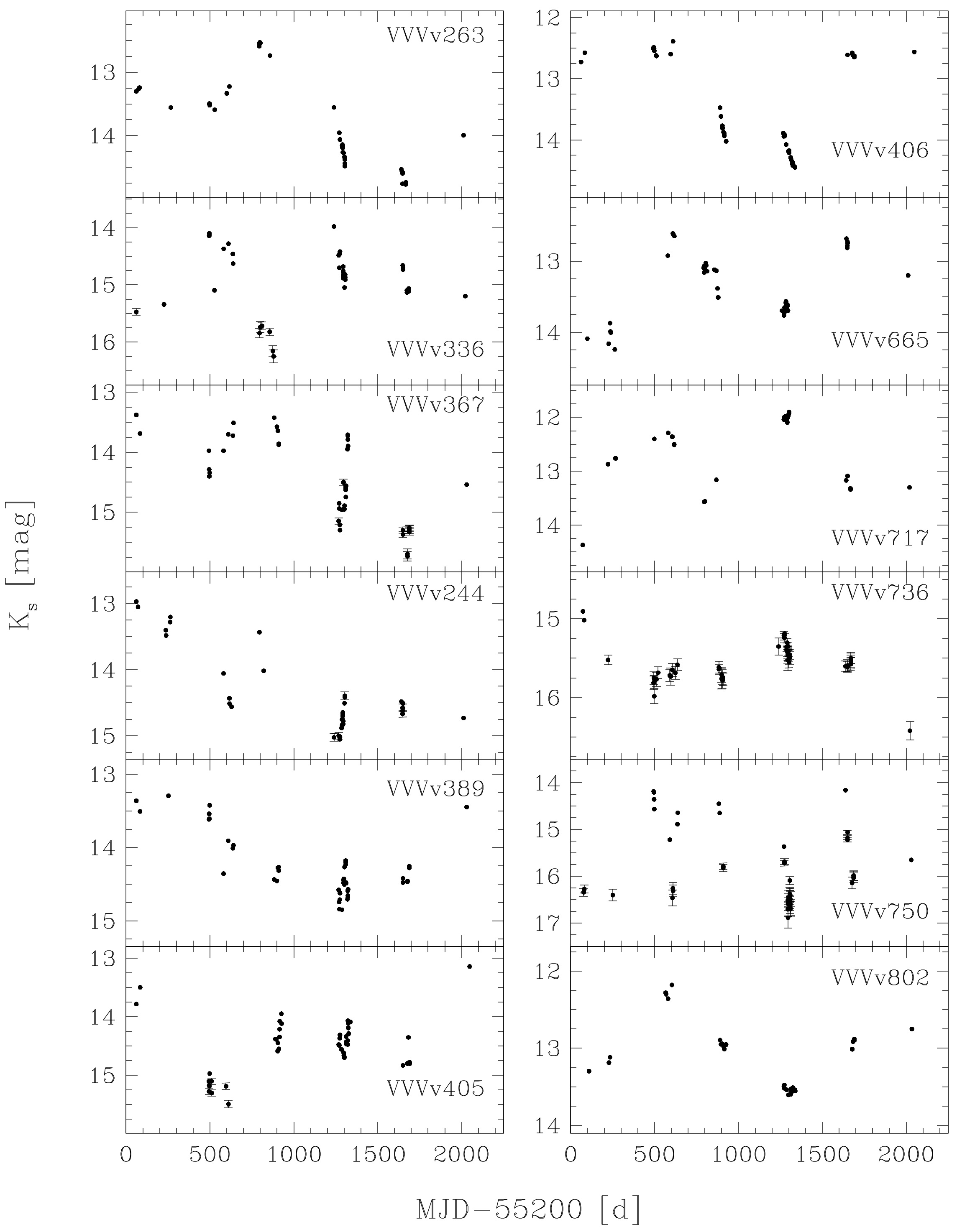}
\figsetgrpnote{Same as Fig.3}
\figsetgrpend

\figsetend

\begin{figure}
\figurenum{3}
\plotone{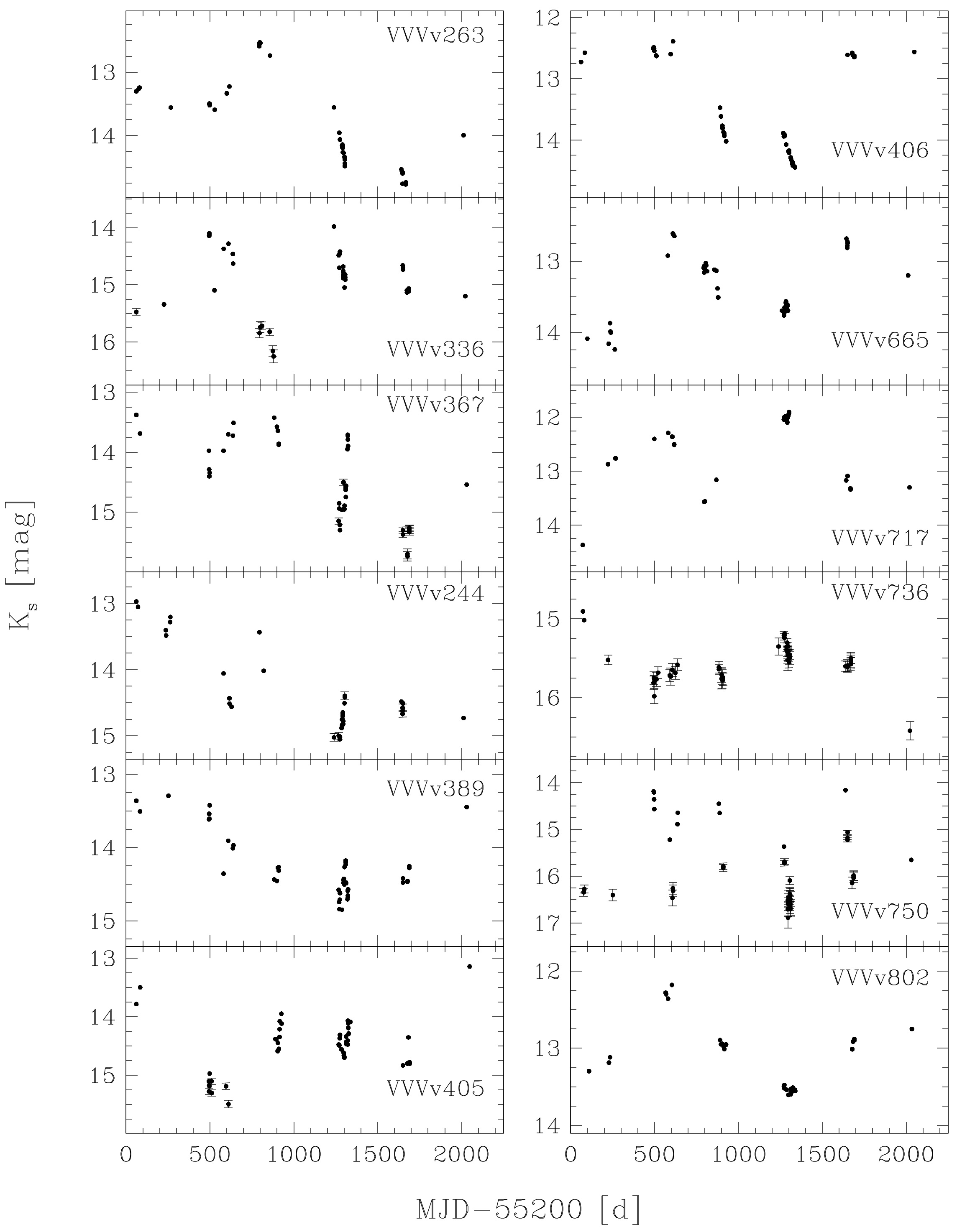}
\caption{Same as Fig.3}
\end{figure}

To carry out such analysis, we used the NASA exo-planet archive
periodogram
tool \footnote{\url{http://exoplanetarchive.ipac.caltech.edu/applications/Periodogram/docs/Algorithms.html}}. This
tool offers various algorithms to process both periodic ( such as
exoplanet or asteroseismic signals) and non-periodic time-series data.
The Plavchan method \citep{plavchan08} is most suited to examine
non-periodic light curve data. This algorithm uses a phase dispersion
minimisation method to detect phased signals by minimising the
residuals to phase-folded light curves with a range of possible
periods. In using the Plavchan algorithm, the user can adjust values
of the number of outlier data points and a phase-smoothing box
size. These parameters were left at default values of 500 and 0.060,
which would allow the liberty to reject all data points (maximum 50
for the data here) if no suitable signal was found. On ingesting the
light curves to the periodogram tool, it attempts to detect and list
phased signals of different frequencies. The signal power is the
magnitude of the coefficients in the frequency domain and the
probability of obtaining the calculated power for a particular period
by chance is given by a p-value. The output is organised to list the
most significant periods first, i.e those with the highest signal
power and least p-value. P-values are in the range between 0 and
1. The interest here is to detect potential phased signals
contributing to the light curve. Therefore, we selected the first
three significant periods, which displayed the least p-value, in most
cases about four to six orders of magnitude smaller than the p-values
of subsequent periods. In other words the first two or three
significant periods (with the highest powers) displayed p-values of
the order of 10$^{-6}$--10$^{-9}$ and the subsequent p-values of the
order of 10$^{-1}$--10$^{-3}$ or larger. The most significant periods
are listed in Table.~4, where, for each VYSO, the columns Per1, Per2
and Per3 show the periods and the columns Power1, Power2, and Power3
are the respective signal powers. If only two periods Per1 and Per2
are listed for a given source, it means that the power spectrum is
relatively clean, lacking further components.
 
From Table.~4 it is evident that the most significant period (Per1) is
of the order of a few hundred days. The median values of Per1, Per2
and Per3 for the 13 sources are 492, 284 and 124 days
respectively. Plavchan method is known to be particularly sensitive in
producing false peaks at integer multiples of the fundamental period
of a real signal, so we compared Per1, Per2, and Per3 for every source
in Table\,4. There are only limited signs of such an artifact here:
some of the periods of the sources VVVv367, VVVv374 and VVVv405 are
close to harmonics.

\begin{figure}
\includegraphics[width=80mm]{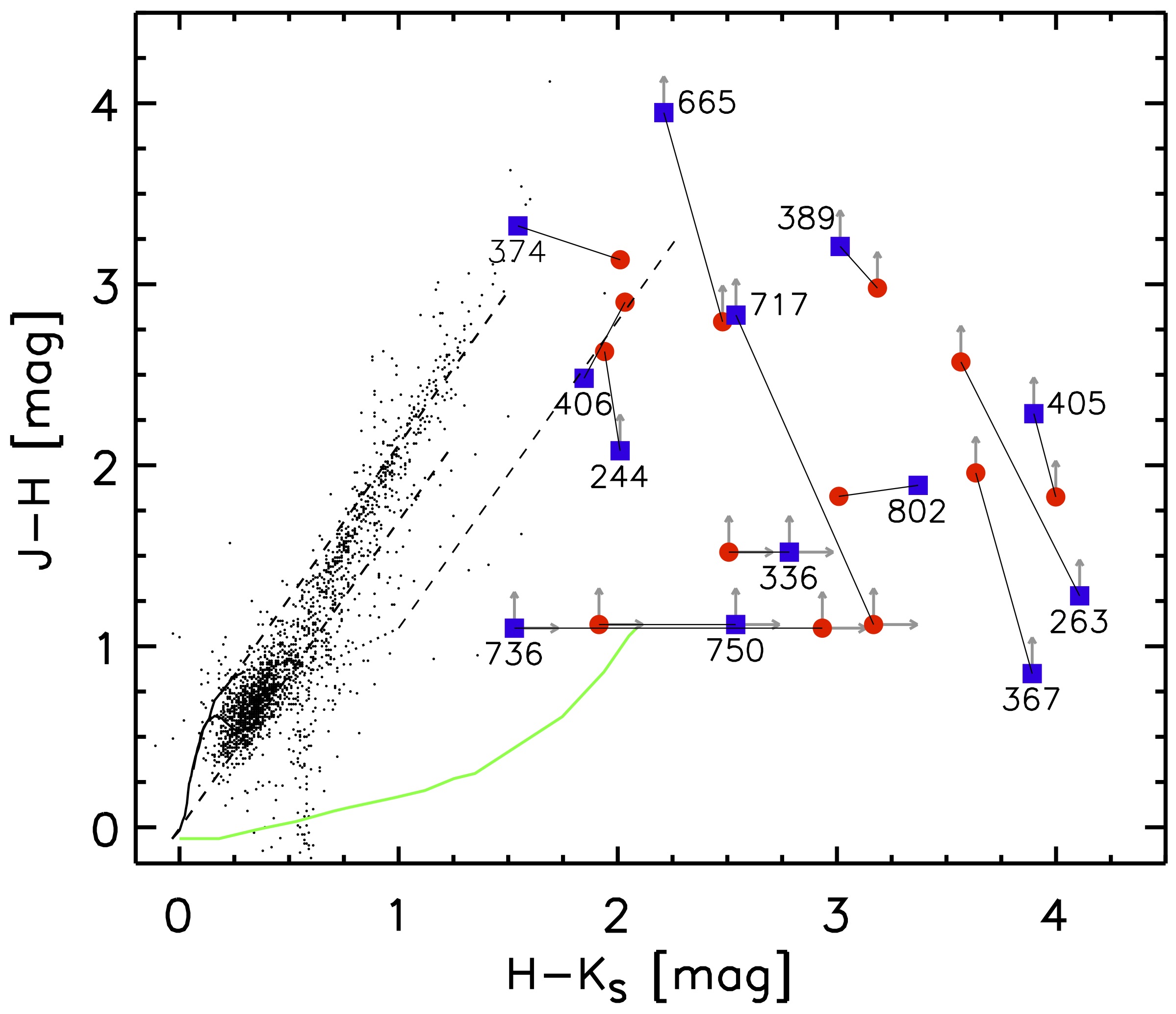}
\caption{Colour-colour diagram. Black dots are the typical
  near-infrared colours of stars in a 6\arcmin $\times$ 6\arcmin field
  centred on the variable sources. The dwarf and giant locus are
  showed in thick black lines. Dotted line show the T-Tauri
  locus. Dashed lines enclosing these curves are reddening vectors
  isolating the reddened groups of those objects. The green curve
  display the HAeBe locus. The red and blue data points display the
  colours of each object at the beginning and end of the light
  curve. }
\end{figure}

\section{Near-infrared colours}

The JHK colour properties of the 13 VYSOs are examined. Multi-band JHK
data is available for two epochs, roughly at the beginning and the end
of the available light curve baseline. In Figure.\,4 we show these two
epochs on the J-H vs H-K colour-colour diagram. The first and last
epoch are represented by red and blue symbols respectively, joined by
a black line. Only three VYSOs are detected in all the three JHK bands
at both epochs. The remaining VYSOs have missing detections, either in
the J, H, or both, in one or both epochs, owing to the very red nature
of these sources. In such cases we have used the VVV sensitivity
limits as an upper limit in computing the colours, and they are
indicated by the arrows. The VYSOs occupy the reddened HAeBe locus
\citep{ladadams92} of the colour-colour space. The HAeBe locus
encompasses intermediate-massive disk (L$_{disk}$=0.1\,L$_{bol}$)
objects with central holes at T$_D \ge$1000\,K. The 13 VYSOs are
reddened by different amounts, as evident from the vertical
distribution of their location parallel to the reddening vector. In
the 5 year timespan, the VYSOs have mostly remained in the HAeBe
space, with some variations in reddening. It may be suggestive of
infrared excess variations as the cause of photometric variability,
although the lack of detections in all the three bands may imply
variable reddening as a possible origin.

\section{Discussion}

In all but one source, the VYSO is offset by $<$5\arcsec\, of the
ATLASGAL peak. The 850\mum peak itself is often associated with a red
diffuse (VVVv736, VVVv406) or point source (VVVv374, VVVv336,
VVVv367). In the exemplary target VVVv374, the red point source
coinciding with the 850\mum peak is found to drive an EGO. This
feature, that the VYSO is located adjacent to an younger and/or more
embedded YSO, may point to a physical association between the two,
with some evolutionary lag, the VYSO having formed first. Numerical
simulations of high-mass star formation \citep[ex][]{krum09} predict
such pairs, the younger companion forming due to fragmentation of a
Jeans unstable massive disk/envelope. In VVVv263, though the 850\mum
emission appears diffuse, a high contrast filamentary dark cloud is
located in the midst of the emission, and the VYSO is centred at the
peak of this infrared filament.

The masses of all 13 VYSOs listed in Table.\,3 are in the range
5-11~\msun, mostly of 7-8~\msun, representing massive YSO
candidates. The distances (from SED models) are in the range
3.5-6.5\,kpc, and the internal core extinction is a hundred to few
hundred visual magnitudes. \citet{carlos16b} have estimated the
distance to five of these VYSOs based on radial velocities and/or
association with star forming regions. The SED modelled distances are
roughly in agreement with those measurements to better than a factor
of two. The models fitted to the source VVVv750 yield an abnormal Av
of 450716\,mag, and it is the only target with a very young age
estimate. This VYSO is associated with one of most circular and
isolated ATLASGAL cores, and its light curve is also the most jittery
(fluctuating), standing out from the rest.

Large-amplitude variability in YSOs can arise due to: episodic
accretion events, varying extinction due to orbital motion of dusty
clumps or a warped disk, or eclipsing binary or multiple
sources. Spots, flares and pulsations can produce only low-amplitude
variability. The non-periodic nature of the VYSO light curves rule out
the possibility of eclipsing near-equal mass objects. High mass star
formation is well-known to produce multiplicity, even at the formation
stages (due to fragmentation), resulting in near-equal mass objects
and/or several low mass objects. The light curves for the sample here
are only sensitive to near-equal mass companions. Varying extinction
if it occurs should reflect in an associated reddening of the
source. The 13 VYSOs are included in the larger sample of
\citet{carlos16a} who have examined the reddening effects and find
that it is not a dominant mechanism to reproduce the high-amplitude
variability. Therefore, the most likely origin of the observed
large-amplitude variability in these intermediate to high-mass VYSOs
must be accretion events. Intermediate resolution near-infrared
spectra for five of the VYSOs (374, 405, 406, 665, 717) have been
discussed by \citet{carlos16b}. All the five sources display shocked
H$_2$ emission at 2.12\mum, irrespective of their light curve
classification. Three of these sources also display Br$\gamma$ in
emission. Both H$_2$ and Br$\gamma$ emission are consistent with the
scenario of ongoing accretion. The case of V723 Carinae as a massive
eruptive variable \citep{tapia15} published during the course of this
study represents another excellent case of the variable phenomenon in
massive young stars.

The analysis of the light curves showed that the most significant
underlying phased signals have periods of the order of a few hundred
days with a median $\sim$500\,days. This timescale and the nature of
variability found here (eruptive, dipper, LPV) are very similar to the
variations of the 6.7GHz class II methanol maser emission in high mass
star forming regions, both in its nature \citep{goedhart04} and in the
timescales \citep{goedhart14}. The class II methanol masers are
thought to be pumped by infrared radiation \citep{cragg05} and found
to originate in accretion disks in high mass star forming regions
\citep{sanna10,sugiyama14}. The similarity in the variability found in
the MME and the IR variability may therefore represent the frequency
of the infrared luminosity variations of the warm disk from accretion
events, which is also responsible for pumping the class II MME
\citep{cragg05}. We searched for MME in the 13 VYSOs and did not find
any methanol masers within a radius of 30\arcsec.
\citet{devilliers15} argue that even though the onset of the 6.7\,GHz
MME is strongly associated with the outflow, and hence accretion
activity, a critical abundance of the methanol molecule is required to
produce the MME. It is therefore likely that the MME sources are
younger and/or embedded in denser regions in comparison with the
evolutionary states of the VYSOs studied here. Nevertheless, the
common origin of both MME and IR-variability is strongly associated
with the accretion phenomenon, in which case, the inferred timescale
of a few hundred days may represent the timescale at which density
enhancements of a spiralling disk feeds the central young star.

\section{Conclusions}

This study has found hitherto unknown near-infrared variability in
intermediate to high-mass young stars by using the VISTA VVV data from
2010-2015. Following a stringent selection criteria to select targets
with $\Delta K \ge 1$mag, 13 VYSOs located at the peak of ATLASGAL
clumps are identified. These sources are characterised by modelling
their 1-850\mum SEDs and by analysing their light-curves with a
phase-dispersion minimisation method.
\begin{itemize}

\item The SED modelling of the 13 VYSO show that their luminosities
  are of $\sim$10$^3$ \lsun, the stellar masses and ages in the range
  of 8-11~\msun and 10$^4$ - 10$^6$\,yrs respectively.

\item The light-curves are not periodic in nature. They can be
  classified as rising, declining or quasi-periodic. Analysis using
  the Plavchan method reveal that the most prominent underlying
  periodic signal would have an average period of $\sim$500days.

\item The high-amplitude variability in young massive stars is
  attributed to episodic accretion events.

\end{itemize}

\floattable
\begin{deluxetable}{lccccccc}
\tablecaption{Source Details \label{table:source}}
\tablewidth{100pt}
\tablenum{1}
\tablecolumns{8}
\tablehead{
\colhead{VYSO} &
\colhead{RA} &
\colhead{Dec} &
\dcolhead{K_s\tablenotemark{a}} &
\dcolhead{\Delta K_s \tablenotemark{b}} &
\dcolhead{ \sigma K_s} \tablenotemark{c} &
\colhead{Light curve} &
\colhead{Association}\\
\colhead{} &
\colhead{deg} &
\colhead{deg} &
\colhead{mag} &
\colhead{mag} &
\colhead{mag} &
\colhead{type\tablenotemark{d}} &
\colhead{}
}
\startdata
VVVv244 & 245.00828 & -51.43392 & 14.4 & 2.08 & 0.59 & \footnotesize{LPV-YSO} & \footnotesize{IRDC}\\
VVVv263 & 245.43405 & -50.34484 & 13.9 & 2.24 & 0.64 & \footnotesize{Eruptive} & \footnotesize{IRDC}\\
VVVv336 & 252.77721 & -45.72340 & 14.9 & 2.27 & 0.51 & \footnotesize{Eruptive} & \footnotesize{IRDC}\\
VVVv367 & 255.12338 & -43.88343 & 14.5 & 2.35 & 0.67 & \footnotesize{Fader} & \footnotesize{IRDC}\\
VVVv374 & 254.64164 & -42.83201 & 11.7 & 2.41 & 0.73 & \footnotesize{Eruptive} &\footnotesize{IRDC, EGO, RMS\tablenotemark{e}}\\
VVVv389 & 255.82157 & -42.43052 & 14.3 & 1.56 & 0.41 & \footnotesize{Fader} & \footnotesize{IRDC}\\
VVVv405 & 257.41092 & -41.64772 & 14.5 & 2.35 & 0.44 & \footnotesize{Dipper} & \footnotesize{HII}\\
VVVv406 & 257.48944 & -41.59691 & 13.5 & 2.06 & 0.78 & \footnotesize{Dipper} & \footnotesize{HII}\\
VVVv665 & 242.49040 & -50.80262 & 13.4 & 1.63 & 0.43 & \footnotesize{Eruptive} & \footnotesize{IRDC}\\
VVVv717 & 249.02318 & -46.67795 & 12.5 & 2.47 & 0.64 & \footnotesize{LPV-YSO} &\footnotesize{IRDC}\\
VVVv736 & 252.73104 & -44.11650 & 15.5 & 1.52 & 0.25 & \footnotesize{Dipper} &\footnotesize{IRDC}\\
VVVv750 & 253.18580 & -43.08889 & 15.8 & 2.73 & 0.82 & \footnotesize{STV} &\footnotesize{IRDC}\\
VVVv802 & 258.54427 & -38.50329 & 13.2 & 1.42 & 0.40 & \footnotesize{LPV-Mira} &\footnotesize{IRDC}\\
\enddata
\tablenotetext{a}{Average magnitudes}
\tablenotetext{b}{The amplitudes for these sources are typically larger than the mean amplitudes for the 816 sample of \cite{carlos16a}, irrespective of the light curve classification.}
\tablenotetext{c}{RMS variation of $\Delta K_s$ as a function of $K_s$ for the given VVV tile}
\tablenotetext{d}{LPV= long periodic variable, STV= short time scale variable}
\tablenotetext{e}{RMS=Red MSX source \citep{lumsden13}}
\end{deluxetable}

\floattable
\begin{deluxetable}{lccccccccccccc}
\tablecaption{Photometric data used for SED fitting \label{table:seddata1} }
\tablewidth{50pt}
\tablenum{2}
\tablecolumns{15}
\tablehead{
\colhead{VYSO} &
\colhead{I1} &
\colhead{I2} &
\colhead{I3} &
\colhead{I4} &
\colhead{W3} &
\colhead{W4} &
\colhead{M1} &
\colhead{PACS70} &
\colhead{PACS160} &
\colhead{SPIRE250} &
\colhead{SPIRE350} &
\colhead{SPIRE500} &
\colhead{AGAL850} \\
\colhead{} &
\colhead{mag} &
\colhead{mag} &
\colhead{mag} &
\colhead{mag} &
\colhead{mag} &
\colhead{mag} &
\colhead{mag} &
\colhead{Jy} &
\colhead{Jy} &
\colhead{Jy} &
\colhead{Jy} &
\colhead{Jy} &
\colhead{Jy}
}
\startdata
VVVv244 &	11.2&	10.0&	8.8&		8.2& 		6.0&		4.1& 		-&		2.4&		10.73&	20.23&	12.88&	 7.42&	 0.84 \\
VVVv263 &	  9.6&	8.1&		6.8&		6.0&		5.40&	2.3\tablenotemark{a}&	1.9&	3.63&	17.23&	-&	 -&	 -&		15.0 \\
VVVv336 &	11.2&	9.7&		8.4&		7.5&		-&	           -&	    4.5&	-&	2.31& 13.47& 15.87& -&	3.84 \\
VVVv367 &	  9.8&	8.3&		7.0&		5.8&		4.6&		1.9&		-&	6.05&	5.75&	7.75&	 -&	 -& 	4.00 \\
VVVv374 &	7.4&	       6.5&	        5.6&	        4.8&	        3.9&	         1.7&   	-&	40.52& 	35.98& 65.92&	 40.79&	 25.93&	3.00 \\
VVVv389 &	10.0&	8.7&		7.9&		7.3&		6.2&		3.6&		-&	0.84&	5.10&	12.99& 15.49& 38.74& 13.42 \\
VVVv405 &	  8.8&	7.4&		6.0&		5.1&		4.4&		-&		2.0&		7.68&	40.31& 89.33& 31.86& 71.33&  7.27 \\
VVVv406 &	10.2&	9.3&		8.5&		7.6&		5.6&		2.5&		-&		3.54&	26.77&	41.01& -& -& 10.78 \\
VVVv665 &      10.0 &	 8.9&		 7.8&		7.0&		6.3 &		 3.8\tablenotemark{a}&		 3.7 & 0.94& 18.46& -& -& -&	1.89 \\
VVVv717 &	10.1&	 8.8&		 7.6&		 6.8&		 6.5&		 4.4\tablenotemark{a} & 4.0 & 0.57 & 2.08 & -& -& -& 2.45 \\
VVVv736 &	10.6&	9.4&		8.0&		7.3&		5.7&		2.9\tablenotemark{a}&		4.5&	67.71&	71.60&	71.79 & 47.11& 28.77& 1.21 \\
VVVv750 &	12.4&	   -&	        9.6&		9.0&		6.3&		2.0&		-&	20.69&	15.56&	21.45& 17.50& 9.80& 1.89 \\
VVVv802 &	8.2&		6.5&		5.4&		4.7&		4.2&		1.7&	-&	31.11&	13.97&	20.31& 14.87& 14.10 & 3.71\\
\enddata
\tablenotetext{a}{not used}
\end{deluxetable}

\floattable
\rotate
\begin{deluxetable*}{LCCCCCCCCCCC}
\tablecaption{SED Fitting results}
\tablecolumns{12}
\tablenum{3}
\tablehead{
\colhead{VYSO} &
\colhead{$M_{\ast}$}&
\colhead{$ A_{V,core}$ } &
\colhead{$A_{V,int}$ } &
\colhead{$\dot{M}_{env}$} &
\colhead{$\dot{M}_{disc}$} &
\colhead{$\log Age$ } &
\colhead{distance}&
\colhead{$\log L_{tot}$} &
\colhead{$\chi^{2}_{best}/N_{data}$} &
\colhead{$N_{fits}$} &
\colhead{$N_{fits2}$\tablenotemark{a}} \\
\colhead{} &
\colhead{$M_{\odot}$} &
\colhead{mag} &
\colhead{mag} &
\colhead{$10^{-5} M_{\odot}$~yr$^{-1}$} &
\colhead{$10^{-6} M_{\odot}$~yr$^{-1}$} &
\colhead{yr} &
\colhead{kpc} &
\colhead{$L_{\odot}$} &
\colhead{} &
\colhead{} &
\colhead{} 
}
\startdata
VVVv244 &  5.9$\pm$2.6 &       192$\pm$4009       & 32$\pm$13  &       7.6$\pm$26.2 &       1.1$\pm$6.0   &  6.2$\pm$0.8 &  5.5$\pm$3.5 &  2.8$\pm$0.7 &  0.05 & 10000 &  1392\\
VVVv263 &  8.5$\pm$2.8 &         97$\pm$595         & 48$\pm$7    &       3.3$\pm$3.5   &       3.8$\pm$27.0 &  6.2$\pm$0.7 &  4.2$\pm$2.6 &  3.5$\pm$0.5 &  0.65 &   684 &    49\\
VVVv336 &  7.3$\pm$2.9 &       206$\pm$2951       & 43$\pm$9    &       0.0$\pm$0.0   &       0.7$\pm$2.8   &  6.4$\pm$0.2 &  5.7$\pm$3.3 &  3.2$\pm$0.6 &  0.09 &  3333 &     8\\
VVVv367 &  8.3$\pm$3.7 &       641$\pm$3857       & 31$\pm$14  &       3.8$\pm$4.2   &       8.7$\pm$34.0 &  5.7$\pm$1.2 &  4.9$\pm$2.9 &  3.3$\pm$0.6 &  0.16 &  1035 &   363\\
VVVv374 &  8.8$\pm$2.7 &       121$\pm$1480       & 15$\pm$4    &       1.6$\pm$7.9   &       0.6$\pm$4.2   &  6.3$\pm$0.2 &  4.4$\pm$2.5 &  3.5$\pm$0.4 &  0.14 &  1659 &   113\\
VVVv389 &  7.3$\pm$2.5 &         67$\pm$1185       & 37$\pm$9    &       0.4$\pm$0.9   &       0.7$\pm$3.1   &  6.4$\pm$0.3 &  5.1$\pm$3.1 &  3.2$\pm$0.5 &  0.38 &  3200 &    75\\
VVVv405 &11.1$\pm$3.3 &       340$\pm$3631       & 45$\pm$6    &       0.0$\pm$0.0   &       1.8$\pm$6.3   &  6.3$\pm$0.2 &  4.4$\pm$2.3 &  3.8$\pm$0.4 &  0.64 &   806 &     0\\
VVVv406 &  5.3$\pm$2.5 &         83$\pm$3391       & 13$\pm$11  &       2.8$\pm$3.2   &       2.8$\pm$8.7   &  4.9$\pm$0.9 &  5.7$\pm$3.6 &  2.6$\pm$0.7 &  0.25 &  3172 &  2812\\
VVVv665 &  7.6$\pm$2.5 &         69$\pm$1056       & 37$\pm$8    &       0.0$\pm$0.2   &       0.7$\pm$2.7   &  6.4$\pm$0.2 &  5.5$\pm$3.2 &  3.3$\pm$0.5 &  0.03 &  3969 &    51\\
VVVv717 &  8.5$\pm$2.8 &         73$\pm$691         & 44$\pm$6    &       0.0$\pm$0.0   &       1.5$\pm$4.1   &  6.3$\pm$0.2 &  5.7$\pm$3.3 &  3.5$\pm$0.5 &  0.49 &  1459 &     0\\
VVVv736 &  7.5$\pm$2.9 &       224$\pm$3415       & 36$\pm$12  &      58.8$\pm$96.1&       0.5$\pm$3.5   &  6.4$\pm$0.3 &  5.9$\pm$3.3 &  3.2$\pm$0.6 &  0.0 &  5515 &   204\\
VVVv750 &  7.4$\pm$3.3 & 450716$\pm$5232901 & 11$\pm$8    &      15.8$\pm$17.0&     19.8$\pm$71.2 &  4.1$\pm$0.7 &  6.2$\pm$3.8 &  2.9$\pm$0.6 &  0.08 &  1807 &  1806\\
VVVv802 &11.5$\pm$3.2 &       160$\pm$2169       & 47$\pm$5    &       0.9$\pm$1.7   &       2.3$\pm$10.0 &  6.3$\pm$0.2 &  3.5$\pm$1.6 &  3.9$\pm$0.4 &  0.29 &  1635 &     5\\
\enddata
\tablenotetext{a}{The number of models with a non-zero envelope accretion rate.}
\end{deluxetable*}

\begin{table}
\tablenum{4}
\caption{Light curve analysis results}\label{tab:sedfits}
\begin{tabular}{@{}lllllll}
\hline
\hline

VYSO & Per1\tablenotemark{a} & Pwr1\tablenotemark{b} & Per2 & Pwr2 & Per3 & Pwr3\\

\hline
VVVv244   &    107.7  & 28.5  &  809.0 &  26.0  &  7.0  &   24.7 \\
VVVv263     &    314.0  & 122.7 &  560.0 &  54.3 & - & - \\   
VVVv336     &    886.3  & 20.8  &  283.5 &  20.1 &  327.1 &  9.5 \\
VVVv367   &    289.7  & 14.8  &  146.9 &  10.7 &   114.0 &  8.5 \\
VVVv374     &    454.0  & 59.1  &  251.3 &  22.4 &   124.3 &  12.1 \\
VVVv389     &    188.5  & 12.2  &  233.1 &  12.2 &   438.7 &  11.5 \\
VVVv405     &    491.5  & 12.4  &  231.1 &  6.5  &   113.4 &  6.5 \\
VVVv406     &    491.9  & 227.1 &  939.0 &  66.6 & - & - \\
VVVv665    &    834.7  & 23.1  &  303.9 &  7.2 & - & - \\
VVVv717    &    851.0  & 177.8 &  277.4 &  68.8 & - & - \\
VVVv736   &    493.1  & 18.8  &  311.5 &  13.3  & 235.6 &  9.5 \\
VVVv750   &    66.2   & 183.9 &  44.4  &  169.2 & - & - \\
VVVv802   &    631.4  & 326.7 &  890.7 &  165.4 & - & - \\
\hline
\end{tabular}
\tablenotetext{a}{Period in days} \tablenotetext{b}{Power of Per1}
\tablecomments{Sources with missing p3 display relatively clean power
  spectrum with two significant periods of probability statistics
  p-value=0}
\end{table}

\hskip 10mm
\acknowledgments


\listofchanges

\end{document}